\documentclass[twocolumn]{article}
\usepackage{graphicx}

\usepackage[a4paper, left=2cm, right=2cm, top=2.5cm, bottom=3cm]{geometry}
 
\usepackage[affil-it]{authblk}
\usepackage[none]{hyphenat}
\usepackage[compact]{titlesec}

\usepackage[backend=biber,
sorting=none,
style=numeric-comp,
giveninits=true,
hyperref,
isbn=false,
terseinits=true]{biblatex}
\DeclareNameAlias{author}{family-given}

\AtEveryBibitem{\clearfield{day}}
\AtEveryBibitem{\clearfield{month}}

\usepackage{amsmath}
\usepackage[separate-uncertainty = true, range-phrase=\,--\,, range-units=single]{siunitx}

\usepackage{float}
\usepackage[font={small}]{caption}
\usepackage{hyperref}

\makeatletter
\renewcommand\AB@affilsepx{\\\protect\raggedright}
\makeatother

\title{Commissioning, characterization and first high dose rate irradiations at a compact X-ray tube for microbeam and minibeam radiation therapy}
\author[1,2,3,4,*]{Christian Petrich}
\author[1,3,4,*]{Johanna Winter}
\author[1,3,5,*]{Anton Dimroth}
\author[6,7]{Thomas Beiser}
\author[1,6,7]{Monika Dehn}
\author[1,4]{Jessica Stolz}
\author[8]{Jacopo Frignani}
\author[1,4]{Stephanie E. Combs}
\author[8,9,10]{Franz Schilling}
\author[5,11]{Ghaleb Natour}
\author[6,7,12]{Kurt Aulenbacher}
\author[1,4]{Thomas E. Schmid}
\author[1,2]{Jan J. Wilkens}
\author[1,3,4]{Stefan Bartzsch}

\affil[1]{Technical University of Munich, TUM School of Medicine and Health, Department of Radiation Oncology, TUM University Hospital, Munich, Germany}
\affil[2]{Technical University of Munich, School of Natural Sciences, Department of Physics, Garching, Germany}
\affil[3]{Heinz Maier-Leibnitz Zentrum (MLZ), Garching, Germany}
\affil[4]{Helmholtz Zentrum München GmbH, German Research Center for Environmental Health, Institute of Radiation Medicine, Neuherberg, Germany}
\affil[5]{Forschungszentrum Jülich GmbH, Institute of Technology and Engineering (ITE), Jülich, Germany}
\affil[6]{Helmholtz-Institute Mainz, Accelerator Design and Integrated Detectors, Mainz, Germany}
\affil[7]{GSI Helmholtzzentrum für Schwerionenforschung GmbH, Darmstadt, Germany}
\affil[8]{Technical University of Munich, TUM School of Medicine and Health, Department of Nuclear Medicine, TUM University Hospital, Munich, Germany}
\affil[9]{Technical University of Munich, Munich Institute of Biomedical Engineering, Garching, Germany}
\affil[10]{German Cancer Consortium (DKTK), Partner Site Munich and German Cancer Research Center (DKFZ), Heidelberg, Germany}
\affil[11]{ISF, Faculty of Mechanical Engineering, RWTH Aachen University, Aachen, Germany}
\affil[12]{Johannes Gutenberg University, Institute for Nuclear Physics, Mainz, Germany}
\affil[*]{Shared first authorship.}
\date{}

\begin{document}

\maketitle
\begin{abstract} 
\textbf{Background:} Minibeam and microbeam radiation therapy promise improved treatment outcomes through reduced normal tissue toxicity at better tumor control rates. The lack of suitable compact radiation sources limits the clinical application of minibeams to superficial tumors and renders it impossible for microbeams. We developed and constructed the first prototype of a compact line-focus X-ray tube (LFXT) with technology potentially suitable for clinical translation of minibeams and microbeams. 

\textbf{Methods:} We give an overview of the commissioning process preceding the first operation, present optical and radiological focal spot characterization methods, and dosimetric measurements. Additionally, we report on first preclinical \emph{in vitro} cell and \emph{in vivo} mouse brain irradiations conducted with the LFXT prototype.

\textbf{Results:} The LFXT was high voltage conditioned up to 300 kV. The focal spot characterization resulted in a strongly eccentric electron distribution with a width of \SI{72.3}{\um}. Dosimetry showed sharp microbeam dose profiles with steep lateral penumbras and a peak-to-valley dose ratio above 10 throughout a \SI{70}{mm} thick PMMA phantom. An open-field dose rate of \SI{4.3}{Gy/s} was measured at an acceleration voltage of \SI{150}{kV} and a beam current of \SI{17.4}{mA} at \SI{150}{mm} distance from the focal spot. \emph{In vitro} and \emph{in vivo} experiments demonstrated the feasibility of the LFXT for minibeam and microbeam applications with field sizes of \SIrange{1.5}{2}{cm}. The mice displayed no observable side effects throughout the follow-up period after whole-brain \SI{260}{\um}-minibeam irradiation.

\textbf{Conclusion:} We successfully constructed and commissioned the first proof-of-concept LFXT prototype. Dosimetric characterizations of the achieved microbeam field showed the superiority of the LFXT compared to conventional X-ray tubes in terms of beam quality. In future developments, the remaining limitations of the prototype will be addressed, paving the way for improved minibeam and first ever microbeam radiation therapy in a clinical setting.
\end{abstract}

\section{Introduction} 

Spatially fractionated radiation therapy (SFRT) is gaining recognition as a promising concept in cancer treatment by breaking with one of the central paradigms of conventional radiation therapy: Instead of delivering a homogeneous dose distribution to the target volume, the radiation field in SFRT is intentionally modulated into low-dose regions (”valleys”) and regions receiving unconventionally high doses (”peaks”). Clinically applied SFRT methods include photon and proton grid and lattice radiation therapy. Several hundred patients with bulky, advanced tumors have been treated with such a dose modulation in the centimeter range \cite{Wu2020, Mayr2024, Mohiuddin19}. 

Preclinical studies indicate that finer dose modulation results in a wider therapeutic window, leading to enhanced tumor control and at the same time reduced side effects such as radiation-induced toxicity \cite{Subramanian2025, Griffin2012}. Thereby, minibeams and microbeams exhibit dose modulations in the millimeter and sub-millimeter / micrometer range, respectively. The wider therapeutic window of microbeams and minibeams compared to a conventional homogeneous dose distribution is attributed to various mechanisms, including anti-tumor immune activation, vascular effects, and bystander responses \cite{Lukas2023, Song2023, Jenkins2024}. Preclinically, microbeam radiation therapy was shown to be highly effective in the treatment of radiation resistance tumors, such as brain, breast, and skin tumors, while minimizing side effects, \textit{e.g.}, in the brain, thorax or abdomen \cite{Fernandez-Palomo2020, Eling2019, Smyth2018}. Recently, Grams et al. \cite{Grams2024} reported first clinical minibeam irradiations of skin tumors using a conventional orthovoltage X-ray tube and a custom-made minibeam collimator fixed to the patient. Patient outcomes indicated symptom relief and positive tumor response. Current limitations of this clinical minibeam radiation therapy approach include very long irradiation times due to low dose rates and the restriction to superficial tumors, mainly due to a smearing of the minibeam pattern with increasing depth in tissue.

A widespread clinical implementation of minibeam and, especially, microbeam radiation therapy is challenging and requires considerable development in compact radiation sources, treatment planning, dosimetry, patient alignment, and standards of dose prescription and quality assurance. The major obstacle remains generating X-ray minibeams and microbeams that retain their sharp dose modulation at several centimeters of tissue depth, without relying on large, non-scalable third-generation synchrotrons such as the European Synchrotron (ESRF) in Grenoble, France, and the Australian Synchrotron in Melbourne, where most preclinical microbeam studies have been performed \cite{Bartzsch2020}. Achieving sharp microbeam dose patterns in patients requires orthovoltage, high dose-rate X-rays emitted from a source with a width comparable to the intended peak width. These characteristics mitigate adverse effects of secondary electron range, geometric blurring, and organ motion. Several compact radiation sources have been proposed for microbeam radiotherapy, including inverse Compton scattering sources \cite{Jacquet2015, Dombrowsky2020}, carbon nanotube-based X-ray tubes \cite{Schreiber2012, Hadsell2014}, and conventional X-ray tubes \cite{Ahmed2024, Bartzsch2016}. However, these technologies have not reached the necessary dose rates and/or photon energies for a widespread clinical application of microbeam or minibeam radiation therapy.

Based on the line-focus X-ray tube (LFXT) concept, we developed and built the first compact X-ray microbeam and minibeam irradiator with technology suitable for clinical application \cite{Bartzsch2017, Winter2020, Matejcek2023}. Here, we report on the commissioning, microbeam dosimetry, and first preclinical minibeam and microbeam irradiations with our proof-of-concept LFXT prototype. 

\section{Methods}

\subsection{Setup and commissioning of the line-focus X-ray tube prototype} 

\begin{figure*}
    \centering
    \includegraphics[width=\textwidth]{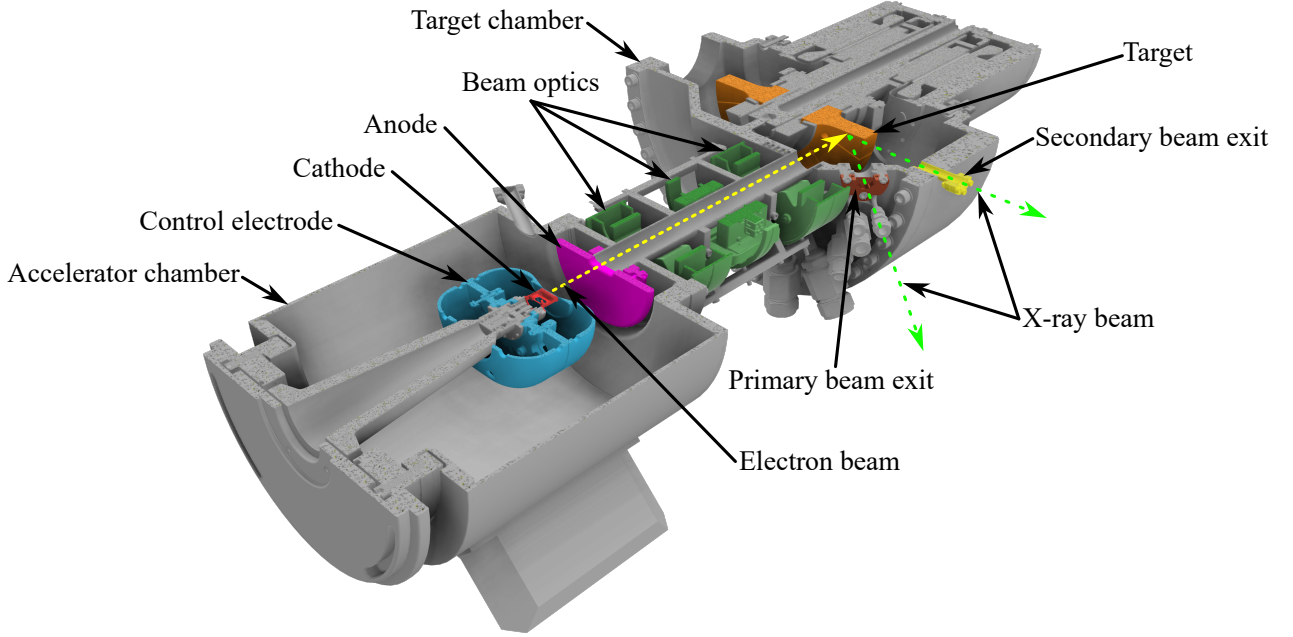}
    \caption{Horizontal section view through the vacuum chamber of the LFXT prototype. The accelerator and target chambers are connected by the electron beam pipe.}
    \label{fig:kammerschnitt}
\end{figure*}

The concept and technology development of the LFXT was described in previous publications \cite{Bartzsch2017,Winter2020,Winter2022,Matejcek2023,Petrich2025}. In brief, our proof-of-concept LFXT prototype resembles conventional X-ray tubes with specific characteristics regarding the electron acceleration and focusing, the target rotation, and the photon beam extraction, as presented in figure \ref{fig:kammerschnitt}. Electrons are emitted from a thermionic cathode, regulated by a control electrode, and accelerated towards a slotted anode by an applied voltage of up to \SI{300}{kV} \cite{Matejcek2023}. The beam pipe is surrounded by two quadrupole focusing magnets and a pair of dipole steering magnets at distances of \SI{150}{mm}, \SI{350}{mm}, and \SI{250}{mm} to the cathode, respectively. The electron beam impinges perpendicularly onto the front surface of the target at a distance of \SI{488}{mm} from the cathode, producing a focal spot with full widths at half maximum (FWHM) of \SI{0.05}{mm}$\times$\SI{22}{mm} according to simulations \cite{Matejcek2023}. The target (Plansee SE, Reutte, Austria) with a diameter of \SI{240}{mm} incorporates a tungsten-rhenium conversion layer and rotates at frequencies of up to \SI{200}{Hz} \cite{Winter2022}. The divergent X-ray beam is extracted at an angle of \SI{45}{\degree} to the electron beam through a primary beam exit window at a distance of \SI{110}{mm} and at an angle of nearly \SI{90}{\degree} to the electron beam through a secondary beam exit window at a distance of 160 mm from the focal spot.

Behind the primary beam exit window, varying apertures, filters, and multislit collimators specifically designed for the LFXT prototype can be placed at a distance of \SI{135}{mm} from the focal spot to shape the X-ray beam into minibeam and microbeam fields. The collimator design accounts for the specific geometry and orientation of the focal spot in relation to the collimator, as presented in \cite{Petrich2025}. Three different collimators were produced with \SI{40}{\um}, \SI{50}{\um}, and \SI{200}{\um}-wide slits, with center-to-center distances of \SI{320}{\um}, \SI{400}{\um}, and \SI{800}{\um}, respectively. The collimators have a square field size of \SI{20}{mm}. A specialized holder allows for precise rotational and translational alignment with the focal spot.

Commissioning of the LFXT prototype began with a two-week bake-out of the vacuum chamber at temperatures up to \SI{200}{\degreeCelsius} to reduce residual gases. Heating tapes were wrapped around the chamber, while ensuring that the heat-sensitive, non-removable quadrupoles remained below \SI{80}{\degreeCelsius}. The electron accelerator was then high-voltage conditioned up to \SI{300}{kV} with a cold cathode to withstand high acceleration voltages without insulation breakdown. This process involved intentionally inducing microscopic vacuum discharges at surface impurities, where locally enhanced electric field gradients caused their abrasion. Four \SI{80}{M\Omega} high-voltage resistors were inserted between the accelerator and the high-voltage power supply during conditioning to limit the current and thereby protect sensitive components from occasional strong discharges. The time-intensive conditioning process was automated using an in-house protocol that adjusted the voltage in response to vacuum levels in the accelerator chamber. For cathode conditioning, the heating current was gradually increased until the cathode temperature reached \SI{1130}{\degreeCelsius}. Subsequently, the high voltage was ramped up while the control electrode retained a potential of \SI{-500}{V} relative to the cathode to prevent uncontrolled electron acceleration. Careful reduction of this potential difference initiated electron acceleration through the anode onto the target, generating first X-rays. 

\subsection{Measurement of the focal spot} 

We employed optical imaging and radiographic pinhole and edge measurements to monitor and characterize the focal spot of the LFXT prototype. The X-ray source was operated during the focal spot characterization experiments with an acceleration voltage of \SI{150}{kV} and an electron beam current in the range of \SIrange{1}{6}{mA}.

\emph{Optical measurement:} High-energy electrons of a few \SI{100}{keV} emit transition radiation in the optical range when passing through material interfaces \cite{Yamamoto2001}. We installed a Raspberry Pi camera to capture this transition radiation from the focal spot. The viewport was oriented at a viewing angle of \SI{67}{\degree} towards the target surface, with the result that the focal spot shape appeared distorted.

\emph{Pinhole measurements:} The pinhole was made from a \SI{1}{mm}-thick tungsten plate with a \SI{25}{\um} diameter hole and centered between the focal spot and a CMOS X-ray detector (S15683-13, Hamamatsu Photonics K.K., Japan). The detector was positioned at a distance of \SI{500}{mm} from the focal spot and had a nominal pixel pitch of \SI{20}{\um}. The geometric magnification was 1 along the short axis of the focal spot and $1/\sqrt{2}$ along the long axis of the focal spot due to the \SI{45}{\degree} angle between the central beam axis and the target surface normal, defined as the axis perpendicular to the target surface originating from the center of the focal spot. Flat- and dark-field image corrections were applied as explained by Petrich et al. \cite{Petrich2025}.

\emph{Edge measurements:} A \SI{2}{mm}-thick tungsten plate with a sharp edge was centered between the focal spot and the CMOS detector. The edge was initially aligned approximately parallel to the long axis of the focal spot to measure the line-spread function (LSF) along the short axis. To optimize rotational alignment, the tungsten edge was rotated around the beam axis in \SI{1}{\degree} steps. Images of each edge position were flat- and dark-field corrected and processed as described by Petrich et al. \cite{Petrich2025}, resulting in individual LSFs. The optimal alignment corresponded to the LSF with the smallest FWHM.

\subsection{Dosimetry} 

After optimizing the electron beam focusing optics, the dose was measured as open-field dose rate, and microbeam and minibeam dose distributions were evaluated within a PMMA phantom and for preclinical experiments.

The open-field dose rate was measured in air at a distance of \SI{131}{mm} from the focal spot. A microDiamond detector was used for dosimetry with a Unidos electrometer (both PTW Freiburg GmbH, Freiburg, Germany) after cross-calibration using a Farmer-type ionization chamber (TM30010, PTW Freiburg GmbH) at a conventional X-ray tube (RS225, Xstrahl, Suwanee, GA, USA) with similar beam quality. Dose rate linearity with respect to beam current was assessed using the microDiamond detector without additional filtering. Acceleration voltages of \SI{135}{kV} and \SI{150}{kV} were applied, while the voltage of the control electrode was gradually reduced, leading to an electron beam current of up to \SI{17.4}{mA}. 

Microbeam and minibeam dose distributions were measured using radiochromic films (Gafchromic EBT3 films, Ashland, Wilmington, DE, USA). Film calibration was performed in the range of \SIrange{0}{10}{Gy} as described previously \cite{Ahmed2024, Bartzsch15}. Peak and valley doses were defined as the means within the central \SI{60}{\%} of the peak and valley regions, with valleys being the regions between peaks defined by their FWHMs. Peak-to-valley dose ratios (PVDRs) were calculated by dividing the average peak dose by the average valley dose.

The peak and valley percentage depth dose curves as well as the PVDR of a microbeam radiation field were evaluated using the collimator with a \SI{50}{\um} slit width and a \SI{400}{\um} center-to-center distance. Measurements were performed in a phantom composed of 14 PMMA plates, each with a thickness of \SI{5}{mm} and a size of $\SI{40}{mm}\times\SI{40}{mm}$, with radiochromic films inserted between the plates. The phantom surface was placed at a distance of \SI{29}{mm} from the collimator, \textit{i.e.}, \SI{163}{mm} from the focal spot. An aperture limited the field size to $\SI{20}{mm}\times\SI{20}{mm}$ at the phantom surface. A \SI{150}{kVp} spectrum was used with a filtering of \SI{0.5}{mm} aluminum, leading to a mean X-ray energy of 53 keV, and the electron beam current was set to \SI{6}{mA}. 

For comparison, a similar measurement was performed at a small-animal irradiator (XenX, Xstrahl) that was in-house transformed into a microbeam irradiation device using the broad focal spot ($\SI{3.55}{mm}\times\SI{2.95}{mm}$), a \SI{225}{kVp} spectrum, and a filtering of \SI{1}{mm} aluminum, resulting in a mean energy of 71 keV of the divergent X-ray beam \cite{Treibel2021, Ahmed2024}. The field size, slit width and distance matched those of the LFXT setup, while the phantom was placed back-to-back to the collimator at a distance of \SI{212}{mm} from the focal spot. 

\subsection{First \emph{in vitro} and \emph{in vivo} experiments} 

Feasibility and proof-of-concept studies were conducted at the LFXT prototype for microbeam and minibeam radiotherapy. Lung carcinoma cells were irradiated with microbeams \emph{in vitro} and Glioblastoma-bearing mice were irradiated with minibeams \emph{in vivo}. The \emph{in vivo} irradiations were conducted as part of a preclinical study, of which we present one representative case as proof of concept.

In the \emph{in vitro} proof-of-concept experiment, established human-derived epithelial non-small cell lung carcinoma cells (A549, ATCC, Manassas, USA) were irradiated at a distance of \SI{150}{mm} from the focal spot using a $\SI{25}{mm}\times\SI{25}{mm}$ microbeam field. The collimator had a slit width of \SI{50}{\um} and a slit center-to-center distance of \SI{400}{\um}. At an acceleration voltage of \SI{135}{kV} without additional filtering and an electron beam current of \SI{8}{mA}, the peak dose amounted to \SI{10}{Gy} with a peak dose rate of \SI{13}{Gy\per\minute}, as measured with radiochromic films. The cells were fixated \SI{30}{min} after irradiation using Paraformaldehyde, stained \SI{24}{hours} later with $\gamma$-H2AX (Merck KGaA, Darmstadt, Germany) for DNA double-strand breaks, counterstained with Hoechst 33342 (Thermo Fisher Scientific, Waltham, USA), and scanned using an inverted microscope (Axio Observer 7, Carl Zeiss Microscopy GmbH, Jena, Germany).

The \emph{in vivo} animal experiment was authorized by the local government authorities (Regierung von Oberbayern, Munich, Germany; license 55.2.2532.Vet-02-20-39) and conducted in compliance with German animal welfare regulations. Six-week-old female C57Bl6/N mice (Charles River Laboratories GmbH, Sulzfeld, Germany) were anesthetized, and 50\,000 GL261 murine glioblastoma multiforme cells (Leibniz Institute DSMZ GmbH, Braunschweig, Germany), were stereotactically inoculated into the right caudate nucleus. Tumor growth was monitored twice per week by T1-weighted MRI using gadolinium-Dota contrast agent (DotaVision, röntgen bender GmbH \& Co. KG, Baden-Baden, Germany). When tumors reached a volume of \SI{15}{mm^3}, total brain minibeam irradiation was performed. 

For irradiation, the anesthetized mouse was placed on a 3D-printed bed designed for brain irradiations with the LFXT prototype and mounted on a six-axis robotic arm (MECA500, Mecademic, Montreal, Canada) for precise positioning. Alignment was verified using a birds-eye camera, a beam’s-eye camera (via a mirror), and a custom Python program displaying the radiation field on a fluorescent screen image.

The minibeam collimator with \SI{200}{\um}-wide slits spaced \SI{800}{\um} apart was used with a squared aperture of \SI{10}{mm} edge length. The isocenter was defined at the center of the brain and mouse bed, located \SI{55}{mm} behind the multislit collimator and \SI{191.5}{mm} from the focal spot. Film dosimetry at the isocenter showed a peak width of \SI{260(4)}{\um}, a center-to-center distance of \SI{1119(3)}{\um}, and a field size of $\SI{13}{mm}\times\SI{13}{mm}$. The dose rate in the peaks amounted to \SI{4.8}{Gy/min} using a \SI{150}{kVp} spectrum with \SI{0.5}{mm} aluminum filtering. This minibeam field behind the primary beam exit window was combined with a broad-beam field behind the secondary X-ray exit window to achieve a PVDR of 10. Thereby, the total peak dose amounted to \SI{150}{Gy} and the valley dose to \SI{15}{Gy} at a depth of \SI{5}{mm}, \textit{i.e.}, at the center of the brain. The isocenter for the broad-beam irradiation was located \SI{266}{mm} from the focal spot. The electron beam current was held at \SI{6}{mA}, yielding a dose rate of \SI{4.8}{Gy/min} in the peak and \SI{0.2}{Gy/min} in the valley behind the primary beam exit window and \SI{0.6}{Gy/min} for the broad beam behind the secondary beam exit window. The mouse was followed up post-treatment with repeated T1-weighted MRI using a RARE sequence on a \SI{7}{T} preclinical scanner (Discovery MR901, Agilent, Santa Clara, CA, USA; Avance III HD electronics, Bruker BioSpin MRI GmbH, Ettlingen, Germany) with a mouse brain 1H surface coil (RAPID Biomedical GmbH, Rimpar, Germany).

\section{Results}
\subsection{Commissioning} 

\begin{figure*}
    \centering
    \includegraphics[width=\textwidth]{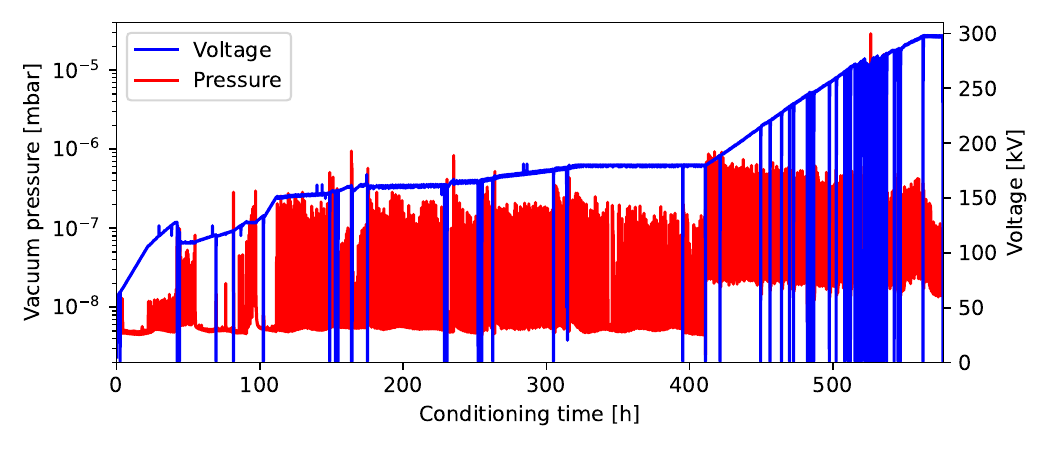}
    \caption{Vacuum pressure and applied voltage during high-voltage conditioning. The red line depicts the vacuum pressure in the accelerator chamber, the blue line depicts the acceleration voltage between cathode and anode. The conditioning time reflects active phases only. Adapted from \protect\cite{Petrich25_diss}.}
    \label{fig:hvconditioning}
\end{figure*}

After bake-out and cooling down, vacuum pressures of \SI{3.9e-9}{mbar} and \SI{2.5e-8}{mbar} were achieved in the accelerator chamber and target chamber of the LFXT prototype, respectively. Figure\,\ref{fig:hvconditioning} shows the vacuum pressure in the accelerator chamber and the corresponding acceleration voltage during high-voltage conditioning of the electron accelerator. The vacuum pressure showed high-frequency pressure spiking, attributed to minor discharges that effectively evaporated surface impurities. The voltage gradient was adapted by improvements to the automatic conditioning protocol. The conditioning was performed up to \SI{300}{kV}.

With the cathode heated and high voltage applied, an X-ray beam was produced by decreasing the potential difference between the control electrode and the cathode. Radiographic images of both open and microbeam fields were then acquired with the CMOS X-ray detector. Due to a malfunction of the rotating target bearing and increasing vacuum levels during irradiation, dosimetry and first experiments were performed at acceleration voltages of only up to \SI{150}{kV} and electron beam powers lower than the design parameters.

\subsection{Measurement of the focal spot} 

\begin{figure*}[t]
    \centering
    \includegraphics[width=\textwidth]{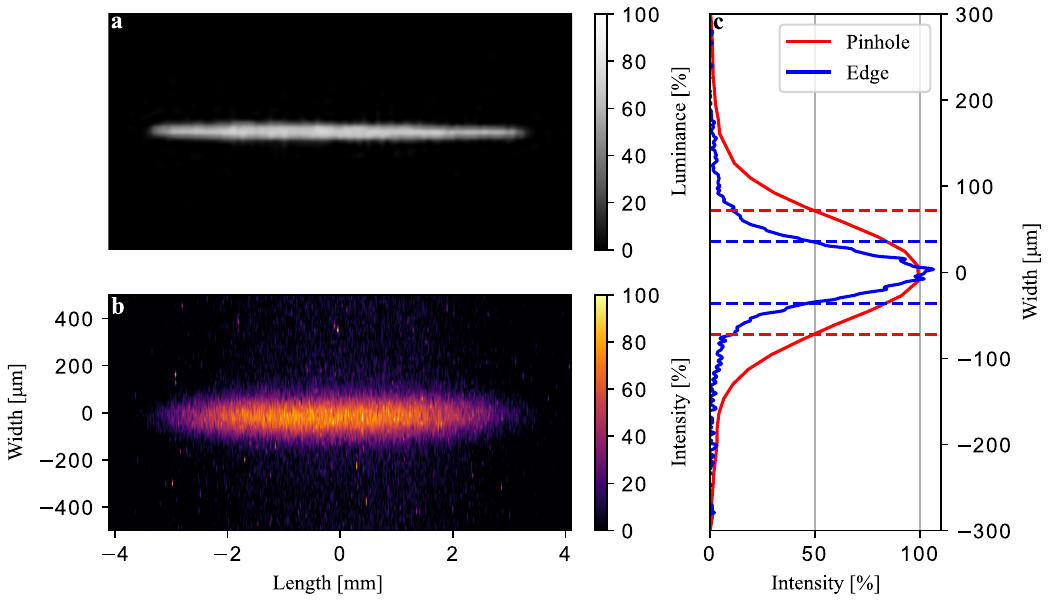}
    \caption{Results of focal spot characterizations using optical and radiological methods. \textbf{a} Normalized luminance of the focal spot visualized with a conventional optical camera. Due to distortions induced by the viewing angle, no absolute dimensions are given. \textbf{b} Radiological intensity distribution in two dimensions (with different scaling) measured using a \SI{25}{\um}-diameter pinhole. \textbf{c} Vertical line-spread functions (LSFs) obtained from the radiological pinhole and edge methods. The LSF resulting from the pinhole measurement exhibits a FWHM of \SI{144.3(8)}{\um}, while the LSF resulting from the edge measurement exhibits a FWHM of \SI{72.3(3)}{\um}, shown as dashed lines. Adapted from  \protect\cite{Petrich25_diss}.}
    \label{fig:FSM_Results}
\end{figure*}

The adjustments of the electron optics were performed stepwise based on the focal spot visualization. Initially, the optical method was used for qualitative visualization while adjusting only the quadrupole magnets. Once preliminary focusing parameters were established, the pinhole and edge methods were applied to quantitatively characterize the achieved focal spot. Figure\,\ref{fig:FSM_Results} presents the results of the three characterization methods. In both the optical image and the radiological image acquired using the \SI{25}{\um} pinhole, the focal spot exhibits the desired strongly eccentric shape, with a length of approximately \SI{6}{mm}. From both radiological methods, the focal spot width was derived by fitting a Gaussian function, resulting in a width of \SI{144.3(8)}{\um} FWHM from the pinhole method and a width of \SI{72.3(3)}{\um} FWHM from the edge method. The difference in measured focal spot width can be explained by the inherent broadening effect of the pinhole on the acquired focal spot projection.

\subsection{Dosimetry} 

\begin{figure*}
    \centering
    \includegraphics[width=\textwidth]{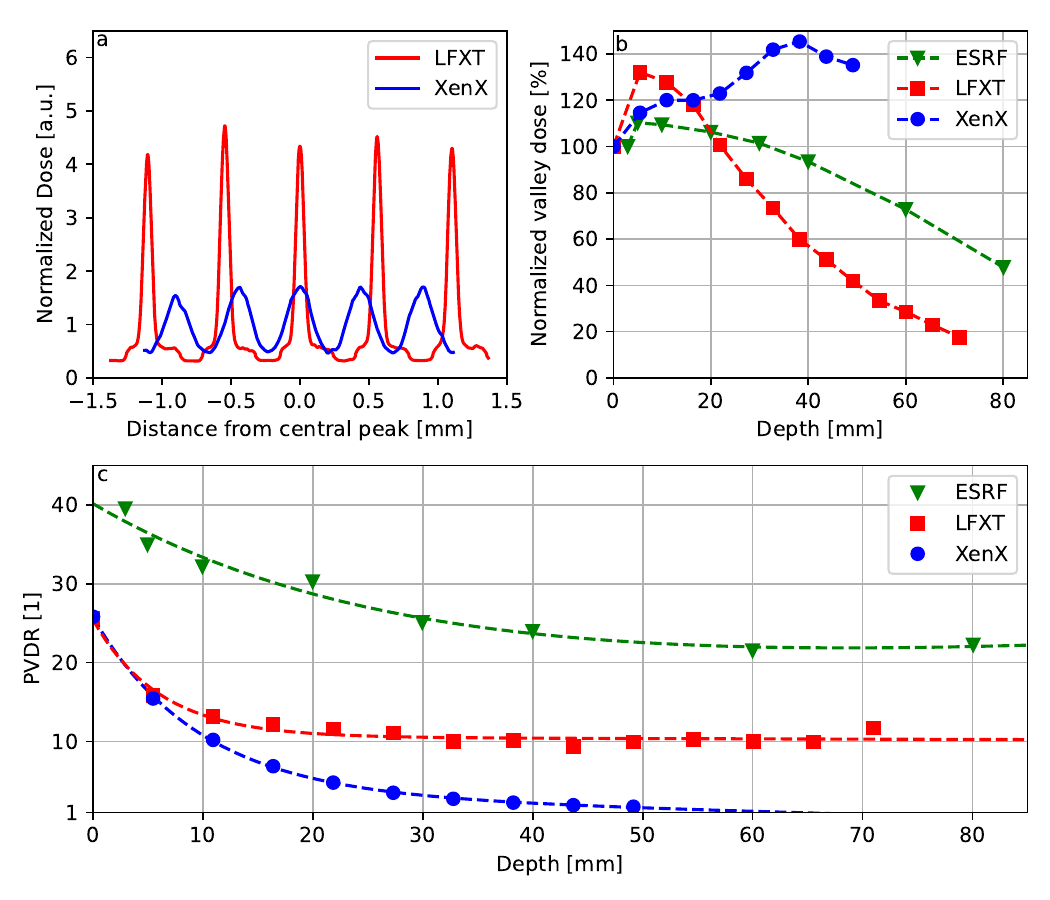}
    \caption{Comparison of microbeam dosimetry at different X-ray sources. \textbf{a} The five central peaks of the microbeam dose profiles at a depth of \SI{22}{mm} measured at the line-focus X-ray tube (LFXT) and a conventional small-animal irradiator (XenX). The profiles are normalized such that their mean values equal 1. \textbf{b} Measured valley depth dose curves for the LFXT, the XenX, and the European Synchrotron Radiation Facility (ESRF) \protect\cite{Bartzsch15}, normalized to their respective entrance dose. The lines visualize a linear interpolation of the measurement points. \textbf{c} Measured peak-to-valley dose ratio (PVDR) in PMMA at the LFXT, the XenX, and the ESRF \protect\cite{Bartzsch15}. Dashed lines indicate the trend of the measurement points fitted to the function $PVDR = exp (a \cdot x + b) + c \cdot x + d$, where $x$ is the depth. The relative standard errors of the mean were below \SI{2}{\%} for the LFXT and XenX measurements and are not shown for clarity. Adapted from \protect\cite{Petrich25_diss}.}
    \label{fig:PVDR_Results}
\end{figure*}

With optimized electron beam focusing, open-field dose rates at the primary beam exit window were obtained from single measurements. The maximum tested electron beam current of \SI{17.4}{mA} at an acceleration voltage of \SI{150}{kV} yielded a dose rate of \SI{4.3}{Gy/s}. A beam current of \SI{12.0}{mA} at an acceleration voltage of \SI{135}{kV} yielded a dose rate of \SI{2.5}{Gy/s}. Both measurements were performed at \SI{150}{mm} from the focal spot. Measurements at lower electron beam currents confirmed a linear relationship between dose rate and beam current, with a steeper slope at the higher acceleration voltage.

Radiochromic film dosimetry showed sharp microbeam dose profiles with steep lateral penumbras throughout the PMMA phantom using the LFXT prototype. Lateral peak penumbras (\SI{80}{\%} to \SI{20}{\%} of the peak-to-valley dose difference) were \SI{27}{\um} at the phantom surface and \SI{41}{\um} at a depth of \SI{71}{mm}. The peak dose rate was \SI{3.3(2)}{Gy/min} at the phantom surface, \SI{2.7(1)}{Gy/min} at \SI{5}{mm}, and \SI{2.0(1)}{Gy/min} at \SI{10}{mm} depth with an electron beam current of \SI{6}{mA}, an acceleration voltage of \SI{150}{kV}, a collimator slit width of \SI{50}{\um}, and a center-to-center spacing of \SI{400}{\um}. The uncertainties mainly resulted from film dosimetry derived from Bartzsch et al. \cite{Bartzsch15}.

Figure \ref{fig:PVDR_Results}a demonstrates considerably sharper dose profiles and higher peak doses with the LFXT prototype compared to the profiles measured at the conventional small-animal irradiator XenX, both obtained at \SI{22}{mm} depth in PMMA. The LFXT yielded a shallower peak depth dose curve than the XenX due to its narrower focal spot, despite having a lower mean energy and greater X-ray beam divergence. As presented in figure \ref{fig:PVDR_Results}b, the valley depth dose curve of the LFXT reached a maximum at \SI{5}{mm} depth due to a build-up of scattered photons, followed by a decrease dominated by absorption and the inverse-squared law. At the XenX, the focal spot being much larger than the collimator slit width resulted in peak smearing and mixing of peak and valley regions, which explains the observed increase in valley dose with depth. For additional comparison, published measurement data from the biomedical beamline ID17 at the ESRF with the same field size of $20\times20$\,mm$^2$ and the preclinical spectrum (mean energy of \SI{105}{keV} \cite{Crosbie2015}) was included \cite{Bartzsch15}. The ESRF achieved shallower peak and valley depth dose curves due to its higher mean energy, which reduced attenuation, and its quasi-parallel beam, which rendered the inverse-squared law negligible. 

As shown in figure \ref{fig:PVDR_Results}c, the PVDR at the LFXT decreased with depth from 25.4 at the phantom surface to 17.0 at \SI{5}{mm} and 13.4 at \SI{10}{mm} depth, leveling off at approximately 10 from \SI{30}{mm} depth to the end of the phantom. PVDRs measured at the XenX were comparable to those at the LFXT up to \SI{5}{mm} depth, but decreased more steeply at greater depths, approaching 1. The better microbeam collimation at the LFXT compared to the XenX resulted from better electron beam collimation, \textit{i.e.}, a smaller focal spot width, which substantially reduced geometric blurring of the microbeam dose profile. It is essential to point out that at the LFXT, the PMMA phantom had a distance of \SI{29}{mm} from the multislit collimator, whereas at the XenX, the phantom was in direct contact with the collimator. Increasing the collimator-to-phantom distance at the XenX would further reduce the PVDR substantially as the geometrical blurring effects of the larger focal spot would be amplified. The relative standard deviation of the mean PVDRs at both the LFXT and the XenX remained below \SI{2}{\%} across all depths. The ESRF, in contrast, produced markedly higher PVDRs than the LFXT, mainly due to its smaller X-ray emitter and higher mean energy.

\subsection{First preclinical experiments} 

\begin{figure*}[t]
    \centering
    \includegraphics[width=\textwidth]{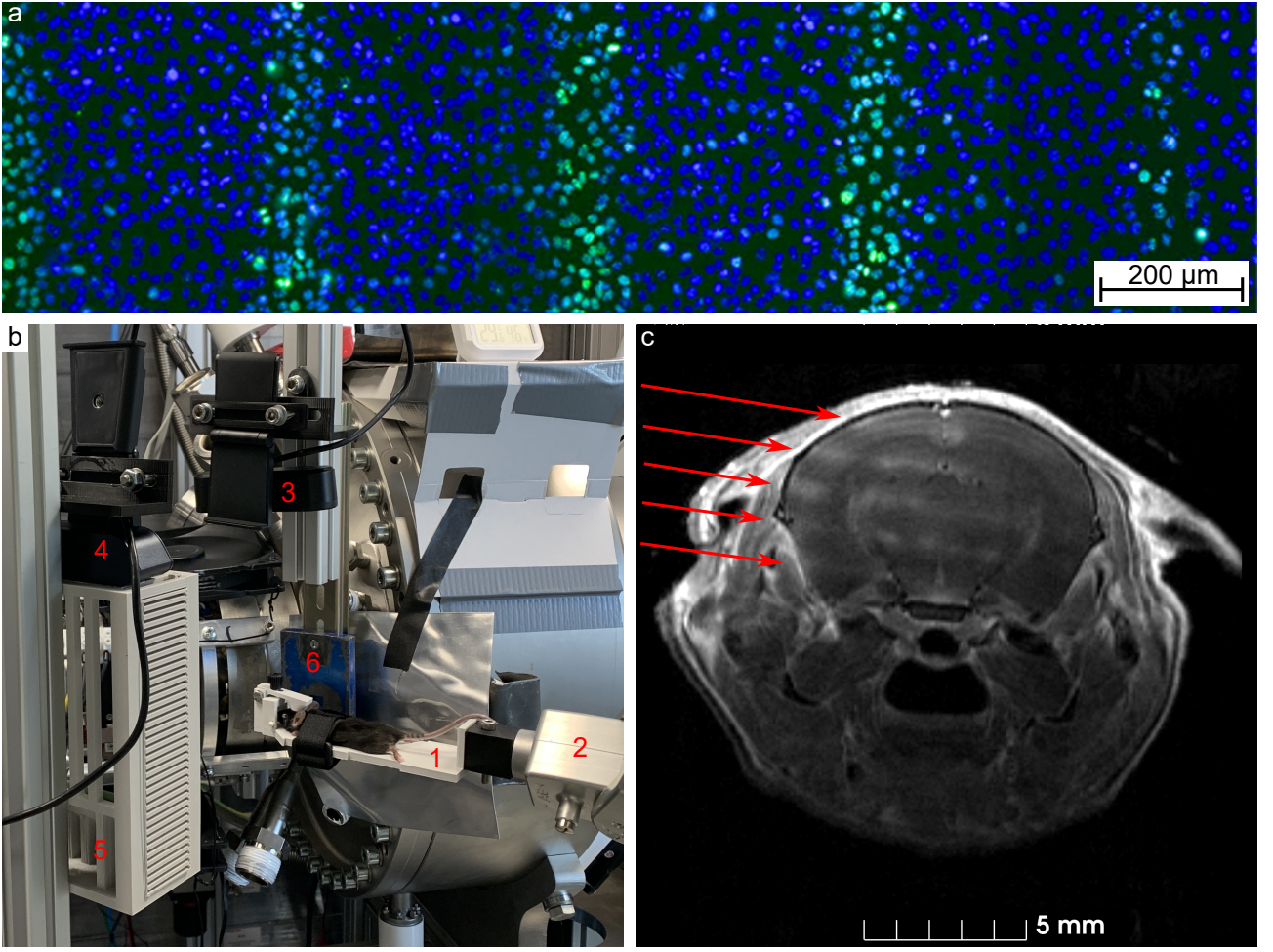}
    \caption{\textbf{a} Near-confluent cell cultures irradiated with microbeams stained against $\gamma$-H2AX (green) and counterstained with Hoechst (blue). \textbf{b} Positioning of a mouse for the irradiation of an orthotopic brain tumor with the mouse bed (1), the robot (2), a birds-eye camera (3), a beams-eye camera (4) with a mirror (5), and collimators (6) in front of the primary beam exit window. c T1-weighted MRI scan of a mouse brain 21 days post irradiation with minibeams visible as bright stripes. Red arrows indicate the entry directions of the individual minibeams.}
    \label{fig:bioResults}
\end{figure*}

Figure \ref{fig:bioResults}a shows a fluorescence microscopy image of human non-small cell lung carcinoma cells, illustrating spatially resolved DNA double-strand breaks \SI{30}{min} after microbeam irradiation. The microbeam irradiation effects are distinctly observable with high-dose regions (peaks) exhibiting pronounced $\gamma$-H2AX foci accumulation, indicating considerable DNA damage. Conversely, the low-dose regions (valleys) show substantially fewer foci, corresponding to reduced DNA damage. This differential response underscores the capability of the LFXT to deliver precise microbeam fields.

Figure \ref{fig:bioResults}b shows the setup used for mouse irradiations at the LFXT prototype. Figure \ref{fig:bioResults}c presents a T1-weighted MR image of an irradiated mouse brain acquired \SI{21}{days} after minibeam irradiation, with the entrance direction of the minibeam field indicated by arrows. The peak regions of the minibeam field are distinctly visible as bright stripes, following intravenous contrast agent injection of gadolinium-DOTA. The intensity of the stripes visibly decreased along the minibeam direction with increasing depth in tissue, indicating a relation between contrast agent uptake and deposited radiation dose. The elevated signal intensities in the minibeam channels after contrast-enhanced T1-weighted MRI are presumably attributed to increased vascular permeability. Throughout the follow-up period of up to \SI{53}{days}, tumor response was observed, and the mouse showed no side effects or neurological dysfunctions attributable to the minibeam irradiation, despite peak doses of \SI{150}{Gy} and valley doses of \SI{15}{Gy}. 

\section{Discussion} 

We demonstrated the implementation of the LFXT as a compact radiation source for microbeam and minibeam radiation therapy. At the LFXT prototype, the microbeam PVDR remained above 10 throughout a \SI{70}{mm}-thick phantom for an entry field size of $20\times20$\,mm$^2$. In contrast, the PVDR values dropped below 3 at a depth of \SIrange{25}{54}{mm} for field sizes of \SIrange{3}{10}{cm} for the first clinical minibeam irradiations \cite{Grams2024}, and at a depth of \SI{31}{mm} for own measurements at a conventional X-ray tube, consistent with previously published data \cite{Ahmed2024}. This rapid decrease in PVDR at conventional X-ray tubes, resulting from their large focal spot size, restricts their application in microbeam and minibeam radiation therapy to superficial tumors. Furthermore, the geometric blurring caused by the large focal spot requires target volumes to be positioned in close proximity to the multislit collimator for conventional X-ray tubes. In contrast, the LFXT maintains sharp dose profiles at several centimeters distance from the collimator, enabling more flexible irradiation positions and treatment of deeper-seated targets. Microbeam experiments at the European Synchrotron Radiation Facility showed higher PVDR values in depth, similar to measurements at the Australian Synchrotron \cite{Dipuglia2019}. The higher PVDR values for the synchrotron experiments resulted mainly from a smaller effective radiation source size (\SI{52}{\um} FWHM \cite{Mittone2020} and \SI{38}{\um} FWHM \cite{Stevenson2010}) and higher mean energies of \SI{105}{keV} \cite{Crosbie2015} and \SI{61}{keV} \cite{Crosbie2013} for the ESRF and the Australian Synchrotron, respectively, compared to \SI{53}{keV} at the LFXT, obtained by own Monte Carlo simulations. Even though synchrotrons provide attractive parameters for preclinical microbeam experiments, they are ineligible for clinical application due to their enormous infrastructural and financial requirements. The asymptotic behavior of the PVDR at the LFXT can be explained by two factors, considering a source size smaller than the collimator slit width. First, geometric magnification affects the peak penumbra and the center-to-center distance equally. Second, photon scattering from peaks into valleys reaches an equilibrium beyond a depth determined by the photon energy. The PVDR achieved at synchrotrons is expected to show the same behavior, but at larger depths due to the higher mean photon energy.

Our study estimates that, at the design parameters of \SI{300}{kV} and \SI{300}{mA}, the LFXT can achieve peak dose rates of \SI{12}{Gy/s} at \SI{5}{mm} depth for a setup comparable to the minibeam mouse irradiations, \textit{i.e.}, at a distance of \SI{200}{mm} from the focal spot. This estimate is extrapolated from the peak dose rate of \SI{4.8}{Gy/min} during the mouse irradiation (\SI{6}{mA}, \SI{150}{kV}, filtering of \SI{0.5}{mm} aluminum), accounting for a factor 3 difference in dose rate efficiency between \SI{300}{kV} and \SI{150}{kV} \cite{Vogt2011}. This estimate is comparable to the previously simulated dose rate of \SI{12}{Gy/s} in a broad beam field at a depth of \SI{2}{mm} at \SI{200}{mm} from the focal spot with a \SI{1}{mm} aluminum filter in the \SI{300}{kVp} spectrum \cite{Winter2023}. In comparison, the first clinical minibeam irradiations were delivered below \SI{1.2}{Gy/min} peak dose rate at \SI{1}{cm} depth \cite{Grams2024}.

At the LFXT, the measured focal spot width of \SI{72.3(3)}{\um} FWHM was approximately \SI{44}{\%} larger than the designed width of \SI{50}{\um}, while the measured focal spot length of \SI{6}{mm} was substantially smaller than the intended \SI{22}{mm}. This shorter focal spot length has no influence on the generated microbeam dose profiles and can be explained by the high potential difference between the control electrode and the cathode to limit the electron beam current, which cuts off the outer portions of the beam. The increased focal spot width can be attributed not only to the ongoing optimization of the electron optics but also to the lower acceleration voltage of \SI{150}{kV} compared to the designed voltage of \SI{300}{kV}, which resulted in a larger geometrical beam emittance and consequently a broader focal spot \cite{Matejcek2023}. As demonstrated, the currently achieved focal spot width is sufficient for generating microbeam and minibeam treatment fields for preclinical research. For operation at higher acceleration voltages and beam currents, the focal spot will be further optimized by putting the steerer magnet into operation.

The \textit{in vitro} and \textit{in vivo} experiments demonstrated the feasibility of the LFXT for preclinical minibeam and microbeam applications. The LFXT concept holds promise for clinical applications of both microbeam and minibeam radiation therapy. The expected increase in dose rate can substantially reduce exposure times, making such treatments more practical and efficient in clinical settings. In contrast to conventional X-ray tubes, the LFXT achieves high dose rates and sharp dose profiles at several centimeter tissue depth, which is essential for effective SFRT. Additionally, the control electrode enables triggering and thereby gating of the X-ray beam in a millisecond range, which improves precision, particularly for moving targets or adaptive treatment strategies. The small size and lower requirements for radiation shielding due to the lower photon energy compared to conventional linear accelerators allow a highly flexible integration into clinical environments.

While the LFXT was successfully commissioned and employed in first preclinical experiments, it remains at a prototype stage, with several challenges still limiting beam power. The development of an improved, second LFXT prototype will focus on optimizing beam quality with a narrower focal spot and higher X-ray energy, improving the design of the rotating target bearing, increasing electron beam current and thus dose rate, and ensuring high reliability of the radiation source. Expanding preclinical studies, including exploiting synergies with immunotherapy, will help to better understand the biological mechanisms underlying microbeam and minibeam radiation therapy and thereby pave the way for clinical translation. The successful implementation of the LFXT into clinical practice could substantially advance cancer treatment by making microbeam and minibeam radiation therapy more accessible and widely applicable in oncology.

\section{Conclusion} 

The first prototype of the LFXT was successfully set up and commissioned as proof of concept. Technical and biological experiments validated its suitability for preclinical microbeam and minibeam radiation therapy. As the LFXT concept is the only suitable compact radiation source that can generate sharp microbeam and minibeam dose distributions at several centimeter depth in tissue with high dose rates, the development and implementation of this LFXT prototype marks a considerable step towards a wide clinical application of microbeam and minibeam radiation therapy.

\section*{Acknowledgments}

This work was supported by the German Research Foundation (Deutsche Forschungsgemeinschaft) [grant number 459947066] and through the Emmy Noether Program [grant number 416790481], the German Federal Ministry of Education and Research (BMBF) through the program GO-Bio initial [grant number 16LW0396 and grant number 16LW0638K], and the Innovation Fund at Forschungszentrum Jülich.

\printbibliography

\end{document}